\documentclass[12pt]{iopart}
\usepackage{graphicx}
\usepackage{subfigure}
\usepackage{dcolumn}
\usepackage{bm}
\usepackage{iopams}

\newcommand{\f}[2]{\frac{#1}{#2}}
\newcommand{\Sum}[2]{\ensuremath{\sum\limits}_{#1}^{#2}}

\begin{document}
\title[A test case for linear theory and violent relaxation]{Linear theory and violent relaxation in long-range systems: a test case}

\author{W. Ettoumi$^{1,2}$ and M.-C. Firpo$^2$}
\address{$^1$ Ecole Normale Sup\'{e}rieure de Cachan, 94235 Cachan, France}
\address{$^2$ Laboratoire de Physique des Plasmas CNRS-Ecole Polytechnique, 91128 Palaiseau cedex, France}
\ead{wahb.ettoumi@ens-cachan.fr}
\ead{marie-christine.firpo@lpp.polytechnique.fr}

\newlength{\textlarg}
\newcommand{\strike}[1]{%
   \settowidth{\textlarg}{#1}
   #1\hspace{-\textlarg}\rule[0.5ex]{\textlarg}{0.5pt}}

\begin{abstract}
In this article, several aspects of the dynamics of a toy model for
long-range Hamiltonian systems are tackled focusing on linearly
unstable unmagnetized (i.e. force-free) cold equilibria states of
the Hamiltonian Mean Field (HMF). For special cases, exact
finite-$N$ linear growth rates have been exhibited, including, in
some spatially inhomogeneous case, finite-$N$ corrections. A random
matrix approach is then proposed to estimate the finite-$N$ growth
rate for some random initial states. Within the continuous, $N
\rightarrow \infty$, approach, the growth rates are finally derived
without restricting to spatially homogeneous cases. Then, these
linear results are used to discuss the large-time nonlinear
evolution. A simple criterion is proposed to measure the ability of
the system to undergo a violent relaxation that transports the mean field modulus in the
vicinity of its equilibrium value within some linear e-folding
times.
\end{abstract}

\submitto{\JPA}


\maketitle

\section{Introduction}

Systems of particles interacting via two-body long-range forces are
well-known to have peculiar equilibrium and non-equilibrium
statistical mechanics (see e.g. Ref. \cite{Elskens} and references
therein). As far as their relaxation properties are concerned, much
progress originated from numerical simulations of the
one-dimensional gravitational system. In 1982, Wright, Miller and
Stein \cite{Wright1982} observed its reluctance to thermalize due to
the existence of quasi-stationary states (QSSs). These observations
were later refined by various authors showing that the relaxation of
the one-dimensional gravitational system usually proceeds through a
rapid approach to a QSS, referred to as violent relaxation, followed
by a very slow drift toward equilibrium
\cite{Reidl1995,Tsuchiya1994_1996}. Such studies initiated a still
very active line of research (see e.g. Refs.
\cite{Yawn2003,Chavanis2005,Joyce2010}) on the intricate interplay
between dynamics, ergodic properties and statistical mechanics in
self-gravitating Hamiltonian systems.

Moreover, in the case where space dimension is larger than one, the
thoroughly investigated gravitational system, as well as the Coulomb
system, combine the difficulties of long-range interaction with a
short-range divergence. This was a motivation to introduce models in
which the potential was truncated to retain only its long-range
components. In addition, the periodic boundary conditions considered
in such models amount to work with a compact space which is
numerically convenient. Various numerical simulations and
theoretical arguments \cite{Elskens,Inagaki1993,Antoni,Sandoz} gave
indications that the corresponding toy models obtained in this way
were sharing purely long-range relaxation characteristics similar to
the original systems.

The Hamiltonian Mean Field (HMF) model \cite{Antoni} derives from
such a truncation procedure as it amounts, in its attractive
ferromagnetic-like form, to the one-dimensional gravitational system
with periodic boundary conditions where only the lowest Fourier mode
is retained. It has become a well-known toy model to address the
intricate relationships between dynamics and statistical mechanics
of long-range interacting systems. It is defined by the following
Hamiltonian
\begin{equation}
\mathcal{H} = \Sum{i=1}{N}\f{{p_i}^2}{2} +
\f{1}{2N}\Sum{i=1}{N}\Sum{j=1}{N}\left[1-\cos\left(\theta_i-\theta_j\right)\right],
\label{eqn:hamiltonian}
\end{equation}
where $N$ is the number of particles, and $\theta_i$ and $p_i$
denote respectively the position and momentum of the
$i^{\mathrm{th}}$ particle. A useful collective quantity to
introduce is the so-called magnetization vector $(M_x,M_y)$ with
\begin{equation}
M_{x} = \f{1}{N} \Sum{i=1}{N} \cos \theta_i \quad\mathrm{and}\quad M_{y} = \f{1}{N} \Sum{i=1}{N} \sin \theta_i
\label{eqn:magnet}
\end{equation}
The average energy per particle $U=\mathcal{H}/N$ reads then
\begin{equation}
U = \Sum{i=1}{N}\f{{p_i}^2}{2N} + \f{1}{2}\left(1-M^2\right),
\end{equation}
where $M\equiv\sqrt{{M_x}^2+{M_y}^2}$ denotes the modulus of the magnetization vector.

Recently, much interest has been devoted to the QSSs which are known
to be responsible for the very slow convergence towards the
statistical mechanics equilibrium predictions. Far from being
difficult to generate, these QSSs naturally emerge in the HMF model
from waterbag initial distributions (see e.g.
\cite{VLatora,Yamaguchi,Leoncini,Firpo,BouchetGupta2010} and the
recent review \cite{Campa}). It is also known that initial waterbag
conditions in momenta, associated to zero or almost zero initial
magnetization, induce the longest lasting QSSs. However, when
lowering towards zero the initial temperature of the particles, it
is possible to exhibit waterbag momenta configurations with
vanishing magnetization in which the magnetization eventually
converges exponentially towards its Boltzmann-Gibbs equilibrium
value. This calls for a linear theory approach.

Linear stability of the HMF model about unmagnetized equilibrium
states has been up to now only studied within the Vlasov framework
\cite{Antoni,Yamaguchi,CampaChavanis2010}, which assumes in particular an infinite
number of particles. Moreover, let alone some very recent publications \cite{CampaChavanis2010}, the linear stability of spatially inhomogeneous, unmagnetized, equilibria has never been considered yet.

The motivation of the present study is then twofold: Firstly and
mostly, one wishes to tackle the linear study of the unmagnetized
cold HMF equilibria, within a finite-$N$, therefore exact,
framework; secondly, the ensuing nonlinear dynamics is briefly
addressed to show that the thermalization of cold unmagnetized HMF
systems finely illustrates Lynden-Bell's concept of violent
relaxation for long-range systems.

In Section \ref{sec_linear}, we shall establish the finite-$N$
framework used for the linear stability derivation. In Section
\ref{section_linearN}, we shall calculate the exact linear growth
rates for two finite-$N$ equilibria, both of zero temperature and
zero magnetization, and compare them to numerical simulations.
Section \ref{sec_random} is dedicated to a random matrix approach
for the calculation of symmetric non-deterministic initial states
growth rates. In Section \ref{section_vlasov}, we eventually derive
the linear theory in the $N \rightarrow \infty$ limit using a fluid
approach derived from the Vlasov equation for a vanishing
temperature. Section \ref{section_violent} ends this study by
discussing the connections between the linear features just derived
and the HMF thermalization properties. The dynamics of the cold
unmagnetized HMF model is proposed as a paradigm of violent
relaxation.

\section{Linear dynamics about cold unmagnetized finite-$N$ equilibria}
\label{sec_linear}

The equations of motion can straightforwardly be written from
Equation (\ref{eqn:hamiltonian}) as
\begin{eqnarray}
\forall \,k\in\left\{1,...,N\right\},\,\left\{
\begin{array}{ll}
\dot{\theta_k} = p_k \\
\dot{p_k}=\f{1}{N}
\Sum{i=1}{N}\sin\left(\theta_i-\theta_k\right)\equiv F_k
\end{array}
\right.
\label{eqn:motion}
\end{eqnarray}
Using Equation (\ref{eqn:magnet}), the force acting on the particle $k$ may be written as
\begin{equation}
F_k=M_y\cos\theta_k-M_x\sin\theta_k.
\end{equation}
Let us consider unmagnetized finite-$N$ equilibria, namely
stationary states of the equations of motion (\ref{eqn:motion}),
with $M_x=M_y=0$. This amounts to have $p^*_k = 0$ and angles
$\theta^*_k$ distributed in such a way that $M_x=M_y=0$. Let us
remark that, since the total momentum $P=\sum_{i=1}^{N} p_i$ is a
constant of motion, the zero momentum equilibrium case considered
here is just the cold (i.e. monokinetic) case for the special choice
$P=0$.

Let us perform the linear stability of this system. In that purpose,
we write $\theta_k = \theta_k^*+\delta \theta_k$ and $p_k = \delta
p_k$ where the asterisk denotes the unperturbed solution. At first
order in $\delta \theta_k$, the force $\delta F_k$ felt by the
particle $k$ verifies
\begin{equation}
\delta F_k =  \f{1}{N}\Biggr[\cos\theta_k^* \,\Sum{i=1}{N}
\cos\theta_i^* \delta\theta_i +
\sin\theta_k^*\,\Sum{i=1}{N}\sin\theta_i^* \delta\theta_i \Biggr].
\end{equation}
This yields the following linear system
\begin{equation}
\left[\begin{array}{c} \left\{\dot{\delta \theta_k}\right\} \\ \left\{\dot{\delta p_k}\right\}\end{array}\right] =
\left[
\begin{array}{cc}
0_N & I_N \\ 
A & 0_N
\end{array}\right] \left[\begin{array}{c}\left\{\delta {\theta_k}\right\} \\ \left\{\delta {p_k}\right\}\end{array}\right]
\label{eqn:system}
\end{equation}
where $I_N$ is the $N\times N$ identity matrix and $A$ a $N\times N$
matrix defined by
\begin{equation}
A_{i,j}=\f{1}{N}\cos(\theta_i^*-\theta_j^*). \label{eqn:defA}
\end{equation}
The stability depends on the eigenvalues $\left\{\lambda_k\right\}$
of the Jacobian matrix in Equation (\ref{eqn:system}). Let us name
it $J$. The eigenvalue problem can now be reduced to the unique
study of $A$ using the following transformation
\begin{eqnarray}
J-\lambda\,I_{2N}&=&\left[
\begin{array}{cc}
-\lambda I_N & 0_N \\ 
A & I_N
\end{array}\right]\cdot \left[
\begin{array}{cc}
I_N & -\f{1}{\lambda} I_N \\ 
0_N & -\lambda I_N + \f{1}{\lambda} A
\end{array}\right]
\end{eqnarray}
provided $\lambda \neq 0$. Hence,
\begin{equation}
\det(J-\lambda I_{2N}) = \det(-\lambda I_N)\det(-\lambda
I_N+\f{1}{\lambda}A) = (-1)^N\det\left(A-\lambda^2 I_N\right).
\end{equation}
In other words, writing $\chi_M$ the characteristic polynomial of $M$, one has
\begin{equation}
\chi_J(\lambda)=(-1)^N\,\chi_A(\lambda^2)
\label{eqn:decomposition}
\end{equation}
This factorization allows us to focus only on the matrix $A$, and
deduce the eigenvalues of the higher-order matrix $J$ by taking the
square root of $A$ ones. We shall now study the case of two
particular finite-$N$ equilibria that advantageously simplify $A$.
\section{Exact finite-$N$ treatment for two special cold force-free equilibria}
\label{section_linearN}
\subsection{The quiet start case}
\begin{figure}[htbp]
    \centering
        \includegraphics[scale=0.4]{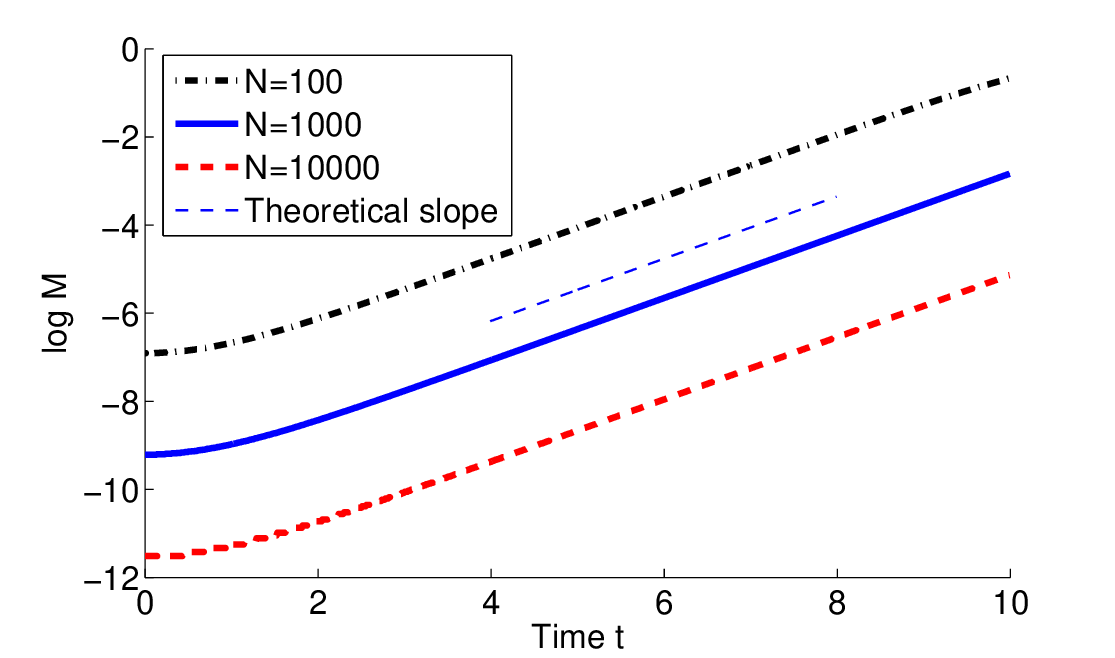}
    \caption{Numerical plot of the magnetization with respect to time. As predicted by Equation (\ref{eqn:qs}), the slope does not depend on the number of particles~$N$, and is in very good agreement with the theoretical value of $1/\sqrt{2}$, plotted in thin dashed line.}
    \label{fig:comparison}
\end{figure}

Having in mind the computational plasma terminology, we define the
so-called "quiet start" configuration as the equilibrium
characterized by an equipartition of the particles on the circle.
It is here formally described by $\forall k$, $\theta_k^* = 2 \pi k
/N$, up to some constant phase, and $p_k^* = 0$. Using this
definition in Equation (\ref{eqn:defA}), the matrix $A$ immediately
reduces to
\begin{equation}
A_{i,j}=\f{1}{N} \cos\left(\f{2 \pi}{N}\left(i-j\right)\right).
\end{equation}
We notice that we can rewrite the coefficients as
\begin{equation}
A_{i,j}=A_{(i-j)\mathrm{ mod }N},
\end{equation}
which makes evident that $A$ is a circulant matrix. Using
\Eref{eqn:decomposition}, it can be easily shown that its
eigenvalues $\left\{\lambda_k ^{2} \right\}$ are expressed as
\begin{equation}
\lambda_k ^{2} = \Sum{j=1}{N} A_j \mathrm{e}^{2 \mathrm{i} j k
\pi/N}. \label{eqn:circ_eigen}
\end{equation}
Moreover, since $A$ is real and symmetric, its eigenvalues are real,
and one can identify the previous equation with its real part
yielding
\begin{eqnarray}
\lambda_k ^{2} &=& \f{1}{N} \Sum{j=1}{N} \cos\left(\f{2 j
\pi}{N}\right)\cos\left(\f{2 j k \pi}{N}\right).
\label{eqn:realeigen}
\end{eqnarray}
This is just
\begin{equation}
\lambda_k ^{2} = \f{1}{2}\left(\delta_{k,1}+\delta_{k,N-1}\right),
\label{eqn:vpA}
\end{equation}
so that $A$ has only one double non-zero eigenvalue equal to $1/2$.
Using \Eref{eqn:decomposition} finally yields the expected
 growth rate $\gamma_{QS}$ coming from the spatially
homogeneous Vlasov linear theory \cite{Antoni} for the cold waterbag
as
\begin{equation}
\gamma_{QS}=\sqrt{\lambda_1 ^{2}}=\f{1}{\sqrt{2}}, \label{eqn:qs}
\end{equation}
with no finite-$N$ correction. In order to test the validity of this
linear study, we performed numerical simulations based on a
fourth-order symplectic integrator \cite{Yoshida}. Starting from a
quiet start configuration, every particle is moved by a uniformly
randomized quantity $\epsilon\ll2\pi/N$. As shown in Figure \ref{fig:comparison}, the behaviour of the system during the early
times shows a very good agreement with the predicted exponential
growth, and does not depend on the number of particles, which only
changes the initial magnetization resulting from the perturbation.
The slow start that one can notice in Figure \ref{fig:comparison} is
due to the fact that the initial random configuration does not
belong to a pure eigenmode, so that the contribution of the other
eigenvalues takes some time to become negligible in front of the
growing one. The same phenomenon can be viewed the instantaneous
growth rates displayed in Figure \ref{fig:general}.

\subsection{Bi-clustered quiet start}
\begin{figure}[htbp]
    \centering
        \includegraphics[scale=0.4]{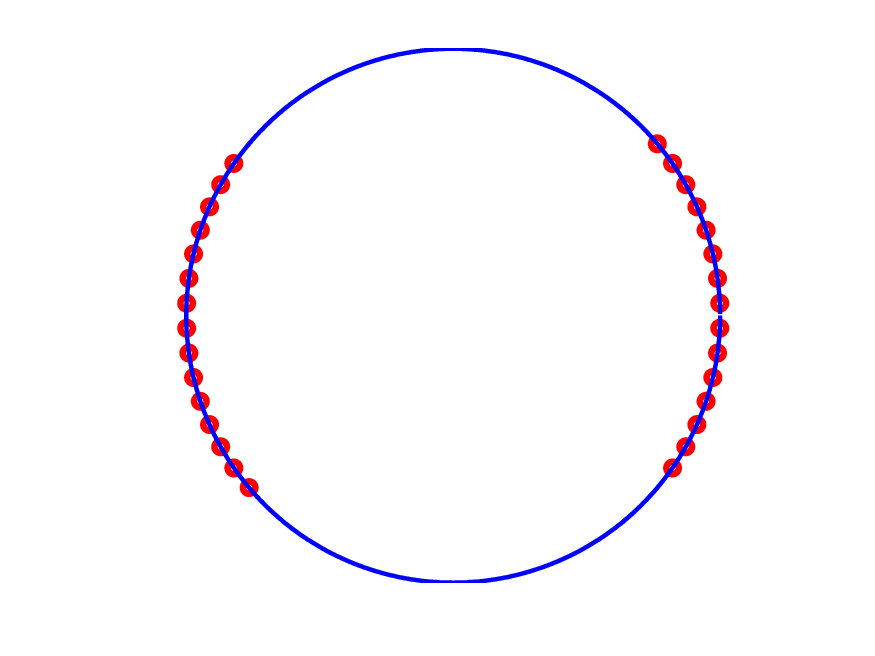}
    \caption{Plot of some finite-$N$ bi-clustered equilibrium where each particle has a zero momentum and faces its symmetric on the circle, providing a zero magnetization.}
    \label{fig:biclusterIC}
\end{figure}

\begin{figure}[htbp]
    \centering
        \includegraphics[scale=0.4]{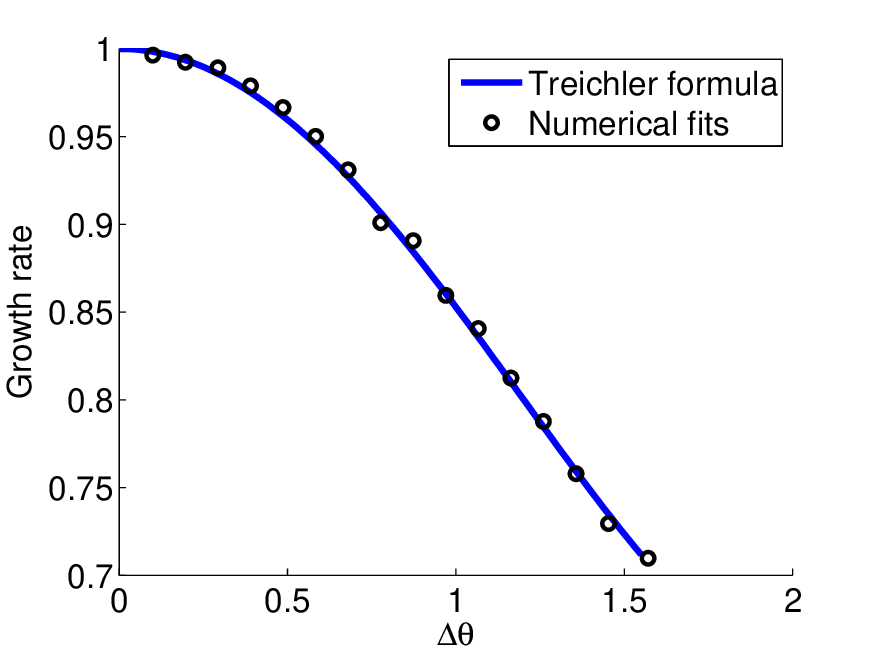}
    \caption{Plot of the growth rate as a function of $\Delta\theta$ for slightly perturbed bicluster initial configurations. The circles correspond to exponential fits from the numerically
integrated magnetization, and the solid curve corresponds to Equation
(\ref{eqn:bcqs}). Each numerical calculation has been performed with
$1000$ particles.}
    \label{fig:accord_BCQS}
\end{figure}

The previous case involved a finite-$N$ analog of a homogeneous
Vlasov force-free equilibrium. We can construct another finite-$N$
equilibrium with zero magnetization by uniformly distributing $N/2$
particles in a cluster of size $\Delta\theta$ centered on a given
position, and by settling the positions of the $N/2$ remaining ones
by rotating the first cluster by $\pi$ with
\begin{eqnarray}
\forall \,k\in\left\{1,...,\f{N}{2}\right\},\,\left\{
\begin{array}{lll}
\theta_k^* = -\Delta \theta + \f{4 k \Delta \theta}{N}\\
\theta_{N/2+k}^* = \theta_k^*+\pi\\
p_k^* = 0 \label{bicluster_sym}
\end{array}
\right.
\end{eqnarray}
Figure \ref{fig:biclusterIC} shows an example of such a bi-cluster
configuration.
The same stability analysis as in the homogeneous quiet start case
can be performed. However, one expects now the growth rate to depend
on $\Delta \theta$, the clusters' size. \Eref{eqn:decomposition} is
still valid, but the matrix $A$ is no longer circulant as it was in
the simple quiet start case. $A$ can be rewritten under the form
\begin{equation}
A=\left[
\begin{array}{cc}
L & -L \\ 
-L & L
\end{array}\right],
\end{equation}
where $L$ is a $N/2 \times N/2$ matrix with coefficients
\begin{equation}
L_{ij}=\f{1}{N} \cos\left(\f{4\Delta\theta}{N}(i-j)\right).
\label{eqn:LetR}
\end{equation}
Therefore, the characteristic polynomial of $A$ reads
\begin{equation}
\det\left(A-\lambda^{2} I_N\right) =
(-2\lambda^{2})^{{N}/{2}}\det\left(L-\f{\lambda^{2}}{2}I_{N/2}\right).
\end{equation}
Or, equivalently,
\begin{equation}
\chi_A(\lambda^{2})=(-2\lambda^{2})^{{N}/{2}}\chi_L\left(\f{\lambda^{2}}{2}\right)
\label{eqn:decomposition2}
\end{equation}
This decomposition allows us to focus on the smaller matrix $L$.
Unfortunately, $L$ is not circulant either, but is a Toeplitz
matrix. Indeed, one can write $L_{ij}=L_{\left|i-j\right|}$. A work
performed by Treichler \cite{Treichler} showed that for a $m\times
m$ Toeplitz matrix generated from the coefficients $t_k=\cos(k
\omega)$, the only two non-zero eigenvalues are
\begin{equation}
\tilde{\nu}_\pm(m,\omega) =
\f{1}{2}\left(m\pm\f{\sin\left(m\omega\right)}{\sin(\omega)}\right).\label{Treichler_result}
\end{equation}
\Eref{Treichler_result} allows us to obtain the eigenvalues
$\left\{\nu_k\right\}$ of the matrix $L$
\begin{equation}
\nu_\pm = \f{1}{N} \tilde{\nu}_\pm\left(\f{N}{2},\f{4\Delta
\theta}{N}\right)=\f{1}{4}\pm\f{\sin(2\Delta\theta)}{2N\sin\left({4\Delta\theta/N}\right)}.\label{Treichler_appl}
\end{equation}
Equations (\ref{eqn:decomposition}), (\ref{eqn:decomposition2}) and
(\ref{Treichler_appl}) give the growth rate as
\begin{equation}
\gamma_{BCQS}=\sqrt{2\nu_+}=\sqrt{\f{1}{2}+\f{\sin(2 \Delta
\theta)}{N \sin(4 \Delta \theta /N)}}, \label{eqn:bcqs}
\end{equation}
that is, in the large $N$ limit,
\begin{equation}
\gamma_{BCQS}=\f{1}{\sqrt{2}} \sqrt{ 1+\f{\mathrm{sin} (2\Delta\theta)}{2\Delta \theta}
}, \label{eqn:bcqs_largeN}
\end{equation}
up to $\mathcal{O}(N^{-2})$ terms. Figure \ref{fig:accord_BCQS}
shows the comparison between the numerically computed growth rates
and the theoretical prediction of Equation (\ref{eqn:bcqs}).
When $\Delta \theta = \pi/2$, particles are uniformly distributed on
the circle as in the previous equilibrium and the growth rate is
exactly $1/\sqrt{2}$ with no finite-$N$ correction. Otherwise, the
growth rate depends on the number of particles but we checked that
the difference with the asymptotic result (\ref{eqn:bcqs_largeN}) is
already very small for $N$ above 10 particles.

\section{Extension to symmetric random initial configurations}
\label{sec_random}

In this section, we will only consider symmetric equilibria prepared
to give $M=0$ in the following way:  we distribute $N/2$ particles
at random on a partition of a given length and put the $N/2$
remaining particles by shifting the random ones by $\pi$. Using the
appropriate indexation of particles (see (\ref{bicluster_sym})), the
calculation of growth rates amounts to determine the largest
eigenvalue of the $N/2\times N/2$ matrix $L$ defined by
\begin{equation}
\forall (i,j)\in{\left\{1,...,\f{N}{2}\right\}}^2,\quad L_{ij}=\cos\left(\theta_i-\theta_j\right),
\end{equation}
where the particle positions $\left\{\theta_i\right\}$ are
distributed according to some $f_0(\theta)$, and where the $1/N$
normalization factor has been voluntarily omitted so that the
coefficients' distribution does not depend on the number of particles.
We used a method based on Random Matrix Theory to calculate the
growth rate's expectation. When the random coefficients verify
$\left\langle L_{ij} \right\rangle_{f_0}=\mu>0$, an extension of
Wigner's law \cite{Furedi} states that the largest eigenvalue is
asymptotically approximated by
\begin{equation}
\nu=\f{2}{N}{\Sum{i,j}{N/2}L_{ij}} + \f{\sigma^2}{\mu} + o\left(\f{1}{\sqrt{N}}\right),
\label{eqn:theoreme}
\end{equation}
where $\sigma^2=\left\langle L_{ij}^2 \right\rangle_{f_0}$. Since
$\left\langle L_{ii}\right\rangle_{f_0}=1$, Theorem $2$ in Reference
\cite{Furedi} states that $\nu$ has a normal distribution of
expectation $1+\left(N/2-1\right)\mu+\sigma^2/\mu$ and bounded
finite variance $2\sigma^2$.
Using Equations (\ref{eqn:decomposition}) and (\ref{eqn:decomposition2}), one finds the mean squared growth rate~as
\begin{equation}
\left\langle\lambda^2\right\rangle=\f{2}{N}\left\langle\nu\right\rangle=\f{2}{N}\left[1+ \left(\f{N}{2}-1\right)\mu+ \f{\sigma^2}{\mu}\right].
\label{eqn:rmtsquared}
\end{equation}
The value of $\lambda^2$ is hence distributed according to a normal
law of variance $8\sigma^2/N^2$ because of the $2/N$ rescaling
factor. The expectation of the growth rate reads then
\begin{equation}
\left\langle\gamma\right\rangle=\mathcal{N}^{-1}\:\int_{0}^{1}
\sqrt{x}\,\mathrm{exp}\left[-\f{N^2}{16\sigma^2}\left(x-\left\langle\lambda^2\right\rangle\right)^2\right]
\rmd x, \label{eqn:rmt}
\end{equation}
where $\mathcal{N}$ is the normalization factor.

This approach simplifies the calculation of the growth rate, since
no more effort in the diagonalization of matrix $A$ has to be done.
However, the most restrictive applicability condition of Equation
(\ref{eqn:rmt}) is the particular symmetry of the initial state,
which allows the use of Equation (\ref{eqn:decomposition2}). The
conditions on the probability distribution of $L_{ij}$ are not as
limiting as the required symmetry of the initial state.

\subsection{Random uniform bi-cluster}

The equilibrium configuration generated by a waterbag distribution
of parameter $\Delta\theta$ yields a state topologically close to
Eqs. (\ref{bicluster_sym}). Therefore, we expect the growth rates
for these random states to be close to the ones given in
\Eref{eqn:bcqs}. However, the latter Equation cannot be the
asymptotic form of Equation ({\ref{eqn:rmt}}) since the highest
eigenvalue fluctuates around $1+\left(N/2-1\right)\mu+\sigma^2/\mu$,
which corresponds to the eigenvalue of the deterministic matrix
$a_{ij}=\mu$ for $i\neq j$, $a_{ii}=1$, that is completely different
from the deterministic matrix $L$ defined by Equation
(\ref{eqn:LetR}).

\begin{figure}[htbp]
    \centering
        \includegraphics[scale=0.4]{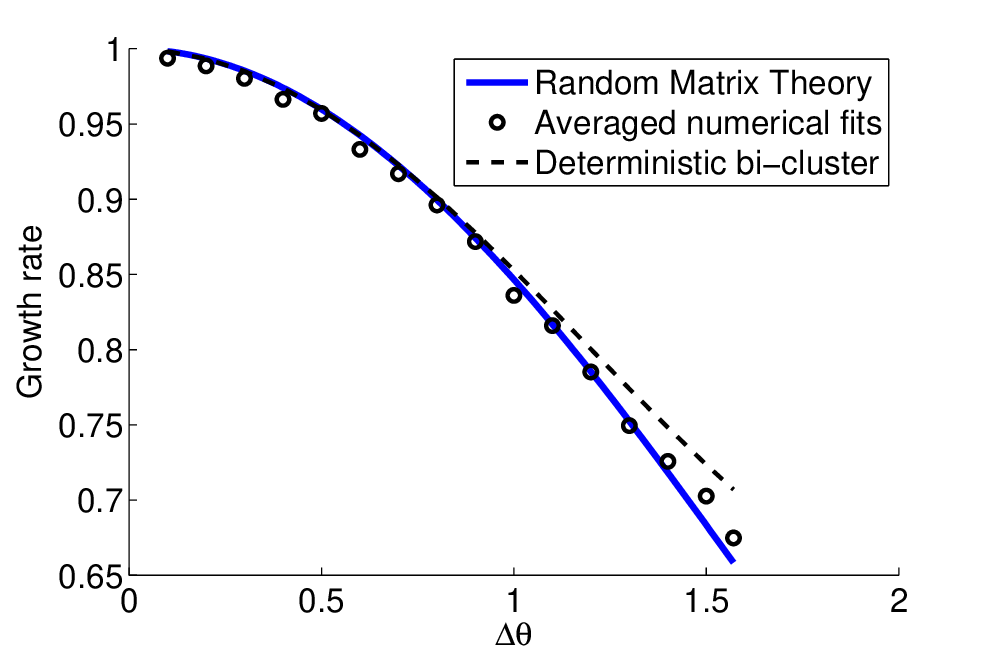}
    \caption{Plot of the growth rates from the uniformly randomized bicluster initial configuration with respect to $\Delta\theta$. The circles correspond to averaged exponential fits from the numerically integrated magnetization, the solid curve corresponds to Equation (\ref{eqn:rmt}). The dashed curve corresponds to Equation (\ref{eqn:bcqs}). Each numerical calculation has been performed with $1000$ particles, and each circle results from the average of $8$ runs. We clearly see a very good agreement with the theoretical prediction.}
    \label{fig:accord_uBCQS}
\end{figure}

One needs to verify the applicability of Equation
(\ref{eqn:theoreme}) before calculating the expectation of the
growth rate with Equation (\ref{eqn:rmt}). $\mu$ and $\sigma^2$ are
accessible through the following formulae
\begin{eqnarray}
\mu = \left\langle\cos\left(\theta_i-\theta_j\right)\right\rangle_{f_0} = \int\!\!\!\int f_0(\theta_i)f_0(\theta_j)\cos\left(\theta_i-\theta_j\right)\rmd \theta_i \rmd \theta_j, \label{eqn:moyenneinteg}\\
\sigma^2 +\mu^2 = \left\langle\cos^2\left(\theta_i-\theta_j\right)\right\rangle_{f_0} = \int\!\!\!\int f_0(\theta_i)f_0(\theta_j)\cos^2\left(\theta_i-\theta_j\right)\rmd \theta_i \rmd \theta_j.
\end{eqnarray}
Writing $\chi(X)$ the characteristic function of the set $X$, the
waterbag distribution of parameter $\Delta\theta$ has the
probability density
\begin{equation}
f_0(\theta)=\f{1}{2\Delta\theta} \chi([-\Delta\theta,\Delta\theta]).
\end{equation}
Hence, we have
\begin{eqnarray}
\mu=\mathrm{sinc}^2\left(\Delta\theta\right), \\
\sigma^2=\f{1}{2}+\f{\sin^2(2 \Delta \theta)}{16 \Delta \theta^2}-\mu^2.
\end{eqnarray}
Clearly, $\mu>0$ for $\Delta\theta < \pi/2$ and $\sigma^2$ is
finite, which means that Equation (\ref{eqn:rmt}) holds. Figure \ref{fig:accord_uBCQS} shows the numerical fits for randomized
bi-clustered initial configurations and the theoretical mean value
given by Equation (\ref{eqn:rmt}).
We also plotted the growth rate (\ref{eqn:bcqs}) corresponding to
the deterministic bicluster to show that, as expected, its behaviour
is close to the expectation of the random one for a wide range of
$\Delta\theta$.

\subsection{Random Gaussian bi-cluster}

\begin{figure}[htbp]
    \centering
        \includegraphics[scale=0.4]{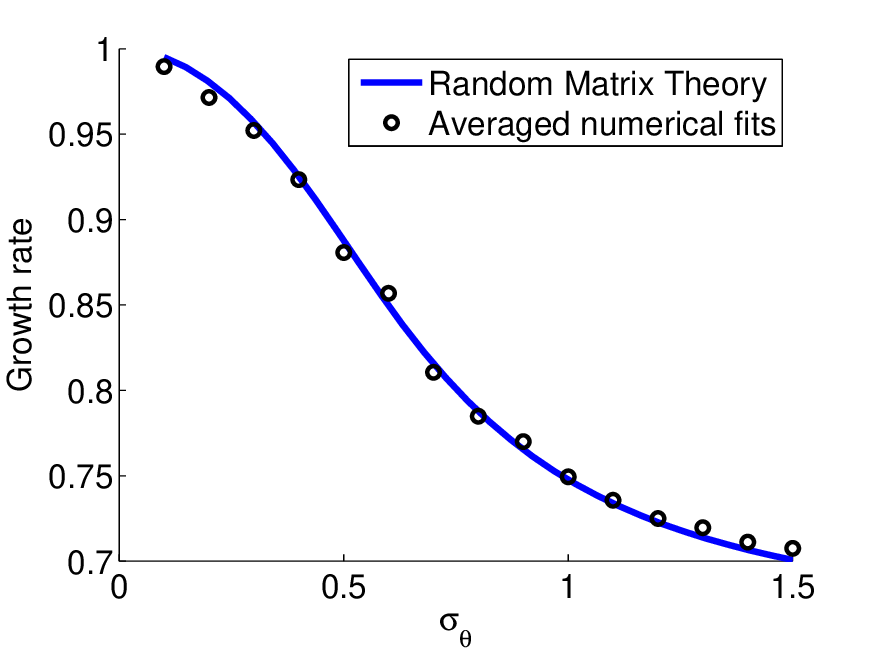}
    \caption{Plot of the growth rates as a function of $\sigma_\theta$ for the random Gaussian bicluster equilibria. The circles correspond to averaged exponential fits from the
numerically integrated magnetization, while the curve corresponds to
Equation (\ref{eqn:rmt}). Each numerical calculation has been
performed with $1000$ particles, and each circle results from the
average of $8$ runs. Here again, the agreement with the random
matrix theory prediction is very good.}
    \label{fig:accord_gBCQS}
\end{figure}

In this subsection we show another example of use of Equation
(\ref{eqn:rmt}) for a more difficult case. The particles are no
longer distributed according to a waterbag density, but with a
Gaussian one. We define $f_0(\theta)$ by
\begin{equation}
f_0(\theta) = \left[\sigma_\theta\sqrt{2\pi}\,\mathrm{erf}\left(\f{\pi}{2 \sigma_\theta \sqrt{2}}\right)\right]^{-1} \mathrm{exp}\left(-\f{\theta^2}{2\sigma_\theta^2}\right).
\end{equation}
The normalization factor has been calculated so that the particles
are distributed on $\left[-\pi/2,\pi/2\right]$ with a standard
deviation $\sigma_\theta^2$. Moreover,
\begin{eqnarray}
\mu = \f{\rme^{-\sigma_\theta^2}}{\mathrm{erf}^2\left(\f{\pi}{2\sigma_\theta\sqrt{2}}\right)} \left[\Re\left\{\mathrm{erf}\left(\f{\pi-2\rmi \sigma_\theta^2}{2\sigma_\theta\sqrt{2}}\right)\right\}\right]^2 > 0\\
\sigma^2=\f{1}{2}+\f{\rme^{-4\sigma_\theta^2}}{8}{\left[\mathrm{erf}^2\left(\f{\pi}{2\sigma_\theta\sqrt{2}}\right)\right]^{-1}}\left[2\Re\left(\mathrm{erf}\left(\f{\pi-4\rmi \sigma_\theta^2}{2\sigma_\theta\sqrt{2}}\right)\right)\right]^2-\mu^2
\end{eqnarray}
As shown on Figure \ref{fig:accord_gBCQS}, the agreement between
the experimental average and the random matrix theory is very good.

In the general case, relaxing the symmetry assumption in the
preparation of the finite-$N$ equilibria yields a vanishing average
value of the matrix elements of $A$ consistently with a vanishing
magnetization. This prevents the application of the Theorem 2 of
\cite{Furedi} and Equation (\ref{eqn:rmt}) no longer holds. To
overtake this difficulty, we now turn to the more usual continuous
approach.

\section{Linear theory within the continuous approach}
\label{section_vlasov}

Writing $f(\theta,p,t)$ the distribution function, one has the
formal Vlasov equation
\begin{equation}
\f{\partial f}{\partial t} + p\f{\partial f}{\partial \theta}+
E(\theta,t) \f{\partial f}{\partial p}=0. \label{eqn:Vlasov}
\end{equation}
One can define the density of particles $n(\theta,t)$ and the mean velocity field $v(\theta,t)$ through
\begin{equation}
n(\theta,t)=\int_{-\infty}^{+\infty} f(\theta,p,t) \rmd p\quad
\mathrm{and} \quad
v(\theta,t)=\f{1}{n(\theta,t)}\int_{-\infty}^{+\infty} p
\,f(\theta,p,t) \rmd p. \label{eqn:n_et_v}
\end{equation}
The HMF force field $E(\theta,t)$ is then given by
\begin{equation}
E(\theta,t)= \int_{-\pi}^{\pi} \sin\left(\alpha-\theta\right)
n(\alpha,t) \rmd \alpha. \label{def_force}
\end{equation}
The early evolution of the cold HMF can be reduced to a fluid
description. The hierarchy of the moments of the Vlasov equation can
be stopped at the first order since the temperature vanishes. By
taking the moment of order zero, one immediately obtains
\begin{equation}
\f{\partial n}{\partial t}+\f{\partial (n v)}{\partial \theta} = 0.
\label{eqn:cold_1}
\end{equation}
Multiplying Equation (\ref{eqn:Vlasov}) by $p$ and integrating over $p$, one has
\begin{equation}
\f{\partial(n v)}{\partial t} + \int_{-\infty}^{+\infty} p^2
\f{\partial f}{\partial \theta} \rmd p -E(\theta,t) n v =0.
\label{eqn:etape1m1}
\end{equation}
Injecting Equation (\ref{eqn:cold_1}) in Equation
(\ref{eqn:etape1m1}) and dividing by $n$ yields
\begin{equation}
\f{\partial v}{\partial t} + v \f{\partial v}{\partial \theta}
-E(\theta,t) = 0, \label{eqn:cold_2_ns}
\end{equation}
in the cold case, for which the mean square of the momentum equals
the square of the mean velocity. Considering the stationary solution
given by some $n_0(\theta)$ yielding a zero magnetization, i.e.
having a zero $m=1$ Fourier component, and $v=0$, one puts $v=\delta
v(\theta)\exp (\mathrm{i}\omega t)$ and $n = n_0(\theta) + \delta
n(\theta) \exp (\mathrm{i}\omega t)$ in Equations (\ref{def_force}),
(\ref{eqn:cold_1}) and (\ref{eqn:cold_2_ns}). Expanding $\delta
v(\theta)$ and $\delta n(\theta)$ in Fourier series, one obtains the
linear system
\begin{eqnarray*}
\rmi \omega \Sum{m}{}\delta n_{m}\exp (\rmi m\theta
)+\rmi\Sum{m}{}\Sum{\ell}{}\left( m+\ell\right) n_{0,m}\delta v_{\ell}\exp \left[ \rmi\left( m+\ell\right) \theta \right]  &=&0, \\
\rmi\omega \Sum{m}{}\delta v_{m}\exp (\rmi m\theta )+\rmi \pi
\Sum{m}{}\delta n_{m}
\left(\delta_{-1,m}\rme^{-\rmi\theta}-\delta_{1,m}\rme^{\rmi\theta}\right)
&=&0.
\end{eqnarray*}
This is
\begin{eqnarray}
\omega \delta n_{k}+k\Sum{m}{}n_{0,m}\delta v_{k-m} &=&0, \\
\rmi\omega \delta v_{\pm1} &=&\pm \f{\pi}{\omega}\delta n_{\pm 1},
\end{eqnarray}
and $\delta v_{m}=0$ for $m\neq \pm 1$. This gives finally
\begin{equation}
\omega ^{2}\delta n_{k}+k\pi\left( n_{0,k-1}\delta n_{1}-n_{0,k+1}\delta n_{-1}\right) =0.
\label{eqn:dispersion}
\end{equation}
The dispersion relation is thus given by $\det M(\omega )=0$, where
$M$ is generally an infinite matrix of elements
\begin{equation}
M_{k\ell}=\omega^2 \delta_{k,\ell} + k\pi
\left(\delta_{\ell,1}n_{0,k-1}-\delta_{\ell,-1}n_{0,k+1}\right),
\end{equation}
with $(k,\ell)\in \mathbb{Z}^2$.

Let us consider the finite-size $2N+1$ square matrix
$\tilde{M}^{(N)}$, which coefficients coincide with $M_{ij}$,
$\forall (i,j) \in \lshad-N,N\rshad^2$. For $N=1$, we have
\begin{equation}
\tilde{M}^{(1)}=\left[
\begin{array}{ccc}
\omega^2+\pi n_{0,0} & 0 & -\pi n_{0,-2} \\ 
0 & \omega^2 & 0 \\
-\pi n_{0,2} & 0 & \omega^2+\pi n_{0,0}
\end{array}\right].
\end{equation}
The condition $\det \tilde{M}=0$ is then fulfilled when
\begin{equation}
\left(\omega^2+\pi n_{0,0}\right)^2=\pi^2 n_{0,2}n_{0,-2},
\label{eqn:omg4}
\end{equation}
but since $n(\theta,t)$ is real, $n_{0,-2}=\overline{n_{0,2}}$ where
the bar denotes the complex conjugate. Therefore, one can take the
square root in Equation (\ref{eqn:omg4}), leading to
\begin{equation}
\omega^2_{\pm}=-\pi n_{0,0} \pm \pi \left|n_{0,2}\right|.
\label{eqn:omg2}
\end{equation}
It is then easy to show by recurrence that  $\forall N, \left|\det
\tilde{M}^{(N+1)}\right|=\omega^4 \left|\det
\tilde{M}^{(N)}\right|$, so that \Eref{eqn:omg2} gives the only
non-vanishing roots to the general dispersion relation.

This allows to obtain the growth rate for the cold unmagnetized HMF
in the infinite $N$ limit as
\begin{equation}
\gamma=\f{\sqrt{1+2 \pi |n_{0,2}|}}{\sqrt{2}}, \label{eqn:V_growth}
\end{equation}
where we used the fact that $n_0(\theta)$ is normalized giving
$n_{0,0}=1/2\pi$.

\begin{figure}[htbp]
    \centering
        \includegraphics[scale=0.4]{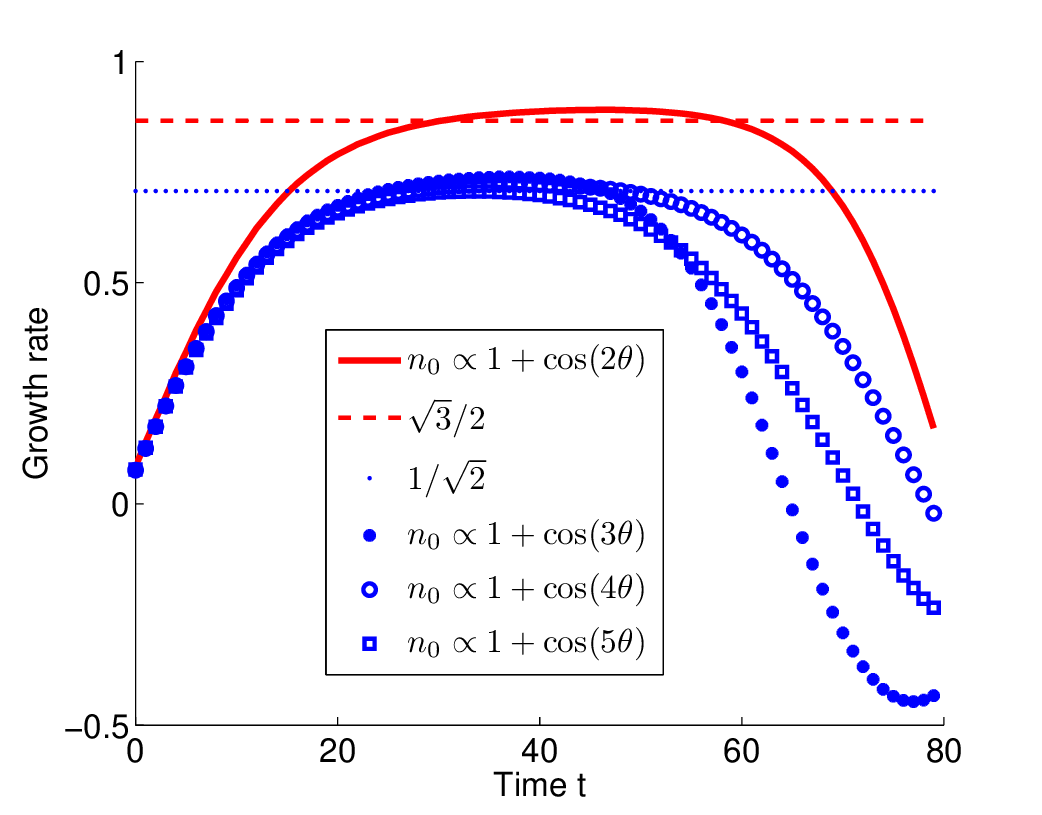}
    \caption{Plot of the numerical growth rate $\dot{M}/M$ with respect to time for different initial equilibria. When the magnetization is almost exponential, $\dot{M}/M$ is practically constant and matches the theoretical value given by Equation (\ref{eqn:V_growth}), represented by the horizontal lines. We clearly see that when the second harmonic of $n_0$ is zero, the growth rates are equal to $1/\sqrt{2}$ as in the homogeneous case. All the runs have been performed using $10^4$ particles.}
    \label{fig:general}
\end{figure}

It is easy to check that the $N\rightarrow \infty$ growth rate
(\ref{eqn:bcqs_largeN}) may be obtained from \Eref{eqn:V_growth} in
the uniform bicluster configuration, for which $2\pi
n_{0,2m}=\mathrm{sinc}(2 m \Delta \theta)$ and $n_{0,2m+1}=0$,
$\forall m \in \mathbb{Z}$. The formula (\ref{eqn:V_growth}) was
successfully tested for a variety of spatially inhomogeneous
equilibria with $M=0$ (See the plots of the instantaneous growth
rates $\dot{M}/M$ with respect to time in Figure \ref{fig:general}).

We shall now briefly discuss the large-time nonlinear features of
the HMF model in the light of Lynden-Bell's picture of violent
relaxation.

\section{Final discussion on the cold HMF case : an example of violent
relaxation}

\label{section_violent}

\subsection{The concept of violent relaxation}

The regularity of the observed luminosity profiles of elliptical
galaxies suggests that they have reached some equilibrium states. As
the two-body, collisional, characteristic timescale is much larger
than the estimated ages of galaxies, the concept of ``violent
relaxation'' was introduced by Lynden-Bell in his famous 1967 paper
\cite{LB1967,White} as a collisionless scenario to account for the
rapid evolution of the galaxies toward quasi-stationary or
quasi-equilibrium states. On the basis of statistical arguments,
Lynden-Bell gave an expression for the coarse grained distribution
function of galaxies in a meta-equilibrium state. Several studies
gave evidence of discrepancies between the numerically obtained
stationary distributions and Lynden-Bell predictions
\cite{Severne1986,Mineau1990} for the self-gravitating system, which
may result from insufficient mixing properties for the application
of this statistical theory. As far as the HMF model is concerned, a
recent application of Lynden-Bell's theory \cite{Chavanis2006} gave
however satisfactory quantitative predictions. In any case, that is
in spite of the controversies related to the strict application of
Lynden-Bell statistics, there appears to be nowadays a common
agreement on the scenario of the relaxation process for $N$-body
long-range systems. It is supposed to be divided in two parts: the
properly speaking violent relaxation part, namely a rapid,
collisionless, evolution of the system towards a quasi-stationary
metastable state, followed by a much slower thermalization phase
towards the state predicted by equilibrium statistical mechanics
(see e.g. Ref. \cite{Gabrielli2010} for recent related results on
generic long-range systems). The HMF model provides a simple, yet
nontrivial, long-range system to consider this issue.

\subsection{Linear instability and violent relaxation in the HMF model}

Extensive numerical studies of the HMF model are now available in
the literature, mostly for two types of initial states: waterbag
initial distribution functions in positions and momenta and
Maxwellian initial distribution functions in momenta with possibly
waterbag initial distribution functions for positions. Long-time
discrepancies between time averaged observables and their ensemble
predictions have been reported only with initial waterbag
distributions in momenta, in connection with the emergence of
so-called quasi-stationary states (see e.g.
\cite{VLatora,Yamaguchi,Pluchino2004}).

Linear theory, that has been up to now mostly formulated within the
Vlasov framework, and for spatially homogeneous states - with the
noticeable exception of very recent extensions to inhomogeneous
states \cite{Barre2010,CampaChavanis2010} -, can be used as a
guideline to discuss relaxation properties in the spirit of
Lynden-Bell's picture. For instance, following the derivation of
Vlasov linear theory given in Ref. \cite{Antoni}, the spatially
homogeneous cases with initial waterbag distributions in momenta are
unstable with growth rates equal to $\sqrt{1/2-3 T}$ where $T$ is
the initial temperature associated to the waterbag. Consequently,
for $T>1/6$ the system becomes linearly stable. This case
corresponds to an energy density $U$ equal to $7/12$
\cite{Antoni,Yamaguchi}. This value happens to coincide with the
energy threshold value above which pathological relaxation
behaviours have been reported for spatially homogeneous, waterbag in
momenta, initial distributions functions.

In the following, we shall discuss the conditions under which
the modulus of the mean-field of the linearly unstable HMF model may
saturate nonlinearly at a value that is close, but yet different, to its ensemble
prediction.

\subsection{Nonlinear saturation of the mean-field}

Let us consider the fluid model introduced in Section
\ref{section_vlasov}. The validity of the zero temperature
approximation is limited to the initial stage of the instability but
this is not critical here as our point is just to establish the
threshold condition at which nonlinear effects come into play to
stop the growth of the initially vanishing magnetization.

In terms of Fourier components, Equation (\ref{eqn:cold_2_ns}) reads
\begin{equation}
\frac{d\delta v_{m}(t)}{dt}+i\sum_{\ell}\ell\delta v_{m-\ell}(t)\delta
v_{\ell}(t) = E_{m}(t), \label{FourierFluid}
\end{equation}
with $E_{m}(t)=0$ for $m\neq \pm 1$, and $E_{1}(t)=E_{-1}^{\ast
}(t)=i\left( M_{x}-iM_{y}\right) /2$. Nonlinear effects are non
longer negligible when the nonlinear contribution balances the other
terms, namely when the
nonlinear mode couplings term balances the linear term. On $m=1$, this yields
\begin{equation}
\frac{d\delta v_{1}(t)}{dt}\sim \sum_{\ell}\ell\delta v_{1-\ell}(t)\delta
v_{\ell}(t)\sim \delta v_{-1}\delta v_{2}  \label{eq_dv1}
\end{equation}
with, on $m=2$,
\begin{equation}
\frac{d\delta v_{2}(t)}{dt}=-i\sum_{\ell}\ell\delta v_{2-\ell}(t)\delta
v_{\ell}(t)\sim -i\delta v_{1}^{2}  \label{eq_dv2}
\end{equation}
Eq. (\ref{eq_dv2}) translates the fact that the $m=2$ mode is
nonlinearly triggered, at about twice the linear growth rate, and
will be the first mode to emerge from the otherwise essentially
$m=1$ instability. The $m=0$ velocity
perturbation remains identically constant. We have%
\[
\delta v_{2}(t)\simeq -i\int^{t}_{0}\delta v_{1}(s)^{2}ds\simeq -i\frac{%
\delta v_{1}^{2}(0)}{2\gamma}\exp \left( 2\gamma t\right)
\]%
so that, replacing this in Eq. (\ref{eq_dv1}), the nonlinear
saturation
takes place when%
\begin{equation}
2\gamma^{2}\sim \delta v_{1}^{2}\simeq \delta v_{1}^{2}(0)\exp
\left( 2\gamma t\right)   \label{NLbalance}
\end{equation}%
together with the linear balance coming from Equation
(\ref{FourierFluid}),
\begin{equation}
\gamma \delta v_{1}\sim E_{1}=\frac{i}{2}\left( M_{x}-iM_{y}\right)
. \label{Lbalance}
\end{equation}%
Eventually the last two orderings (\ref{NLbalance}) and
(\ref{Lbalance}) give the order of the modulus of the magnetization
at the nonlinear saturation, $M _\mathrm{sat}$, as
\begin{equation}
 M _\mathrm{sat}\sim 2^{3/2}\gamma^{2}.
\label{NLthreshold}
\end{equation}
This threshold is qualitative. In order to obtain a more
quantitatively valid estimate, which is not an easy task at all, one
should have to take into account, in particular, the fact that the
growth rate does not remain equal to its linear value up to
nonlinear saturation.

Let us use the common plasma physics terminology for wave-particle
interaction and introduce the trapping time of the particles in the
mean-field potential well, namely the inverse of the bounce
frequency $\omega_b = \sqrt{M_\mathrm{sat}}$ at the nonlinear
saturation. This is just the characteristic timescale for particles
oscillating in the HMF potential well, and does not depend on $N$.
This is a so-called nonlinear timescale as the mean-field is
supposed to be initially vanishingly small. The linear timescale is
obviously given by the e-folding time, namely by the inverse of the
linear growth rate $\gamma$. Then Equation (\ref{NLthreshold})
translates the fact that nonlinear saturation takes place when both
timescales balance, namely for $\gamma \sim \omega_b$. As a
consequence of Equation (\ref{NLthreshold}), one expects nonlinear
saturation to take place close to equilibrium predictions provided
that the linear growth rate is of the order of the ensemble average
of the nonlinear frequency, namely provided that
\begin{equation}
\gamma \sim \langle \sqrt{M} \rangle_{\mu}. \label{condition}
\end{equation}
On the contrary, if the linear growth rate $\gamma$ is small enough
so that $\gamma \ll \left\langle \sqrt{M}\right\rangle_\mu$, the
value of the nonlinear magnetization threshold will be significantly
below the value predicted by equilibrium statistical mechanics.

\subsection{Application to the cold beam case}

The cold case, on which we have just focused, illustrates well this
scenario. Actually, large-time simulations show that the modulus of
the magnetization, for instance, quickly converges towards a
saturated state with a value that is close to its ensemble
prediction.

Let us check that the condition (\ref{condition}) is indeed
satisfied for the cold unmagnetized HMF. This corresponds to an
energy density $U=1/2$. According to equilibrium statistical
dynamics \cite{Rocha}, this gives $\langle M \rangle _{\mu}=\langle
M \rangle _{c}=1/\sqrt{\beta}\equiv \sqrt{T}$ where the inverse of
the temperature is given implicitly by
$\rm{I}_{1}(\sqrt{\beta})/\rm{I}_{0}(\sqrt{\beta})=1/\sqrt{\beta}$.
Numerically, the ensemble average of the magnetization for the cold
HMF is then $\langle M \rangle _{c}\simeq0.62$ and the equilibrium
temperature is $T\simeq0.39$. Since $\gamma=1/\sqrt{2}$, one effectively has $\gamma \sim
\langle \omega_b \rangle _{\mu}$, where we have used ensemble
equivalence.

It is interesting to note that similar observations have been
reported for another long-range, mean-field, wave-particle
Hamiltonian model starting with a cold beam of particles \cite{FLA}.

\subsection{The relaxation process}

\begin{figure}[htbp]
\begin{center}
  \subfigure{\includegraphics[scale=0.4]{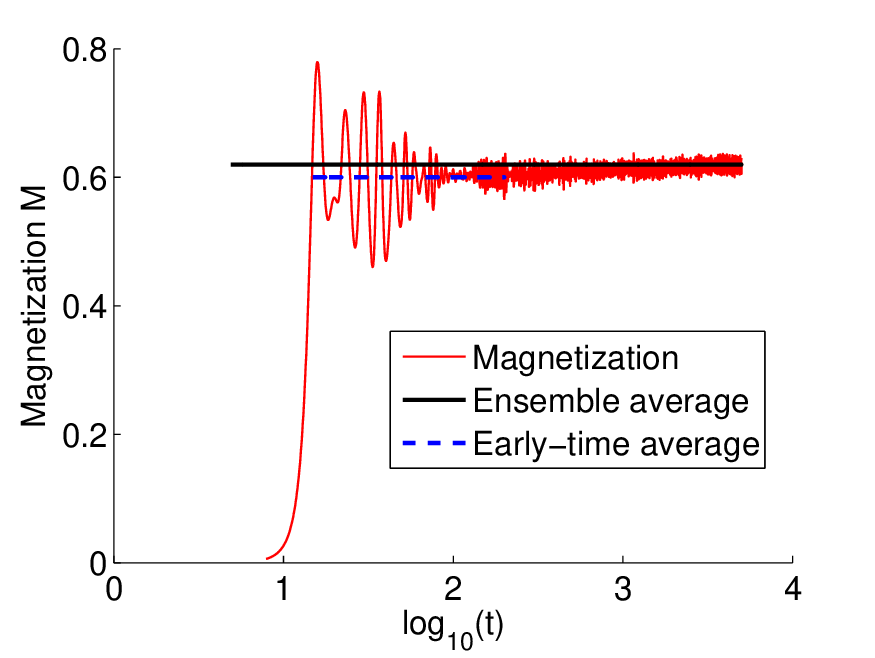}}
  \subfigure{\includegraphics[scale=0.4]{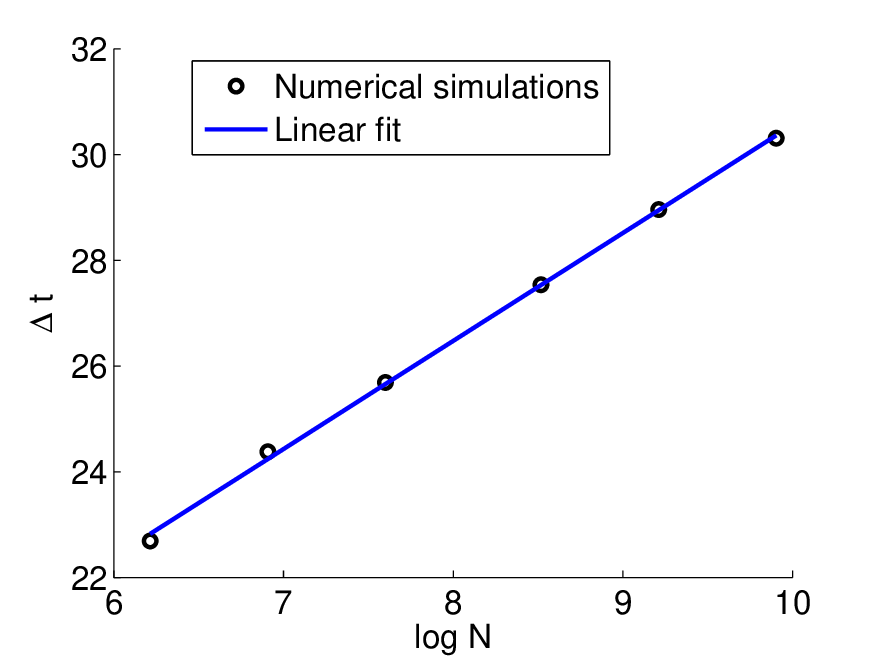}}\\
\end{center}
\caption{(a) Time evolution of the magnetization starting from a
gently perturbed homogeneous quiet start of the cold HMF. (b) $N$
dependency of $\Delta t$, defined as the time needed for the
magnetization to first reach the value 0.7 that roughly corresponds
to nonlinear saturation.} \label{fig:monitoringM}
\end{figure}

Figure \ref{fig:monitoringM}a shows that the cold unmagnetized HMF
does actually experience an initial violent relaxation phase. The
figure makes however apparent that this phase is followed by a much
slower thermalization phase that is needed for phase space sweeping
and complete convergence towards equilibrium statistical predictions.
Actually, in the early nonlinear saturation, the magnetization
oscillates about $0.60$, which is below the ensemble equilibrium value. The
drift towards the equilibrium statistical predictions takes place on
a much longer timescale than the violent relaxation timescale.

More precisely, putting $\delta M_{0}$ the initial infinitesimal
perturbation of the modulus of the magnetization, that may contain
some $N$-dependence for the finite-$N$ HMF model, the time $\Delta
t$ needed to reach the nonlinear saturation threshold is of the
order of $\gamma^{-1} \ln (\gamma^{2}/\delta M_{0})$. Assuming that $\delta M_0 \propto N^{-1/2}$,
one gets $\Delta t \propto \log N$, which is confirmed by the plot on Figure \ref{fig:monitoringM}b.

The ensuing timescale needed for complete thermalization of the system has a
much stronger $N$-dependence, as it is basically a binary
collisional timescale that diverges with $N$. Therefore, in the
Vlasov $N \rightarrow \infty$ limit, the system would have been
trapped in the QSS corresponding to nonlinear saturation. The
specific thermalization process will be more closely examined in a
forthcoming study.

\bibliographystyle{unsrt}

\begin{thebibliography}{10}

\bibitem{Elskens} Elskens Y and Antoni M 1997, \textit{Phys. Rev. E} \textbf{55}  6575–6581

\bibitem{Wright1982} Wright H L \textit{et al} 1982, \textit{Astrophys. Space Sci.} \textbf{84} 421

\bibitem{Reidl1995} Reidl C J Jr. and Miller B N 1995, \textit{Phys. Rev. E} \textbf{51} 884–888

\bibitem{Tsuchiya1994_1996} Tsuchiya T \textit{et al} 1994, \textit{Phys. Rev. E} \textbf{50} 2607–2615; \textit{ibid.} 1996, \textit{Phys. Rev. E} \textbf{53} 2210–2216

\bibitem{Yawn2003} Yawn K R and Miller B N 2003, \textit{Phys. Rev. E} \textbf{68} 056120

\bibitem{Chavanis2005} Chavanis P-H 2005, \textit{Astron. Astrophys.} \textbf{432} 117

\bibitem{Joyce2010} Joyce M and Worrakitpoonpon T 2010, \textit{J. Stat. Mech.: Theory Exp.} P10012

\bibitem{Inagaki1993} Inagaki S and Konishi T 1993, \textit{Publ. Astron. Soc. Jpn.} \textbf{45} 733

\bibitem{Antoni} Antoni M and Ruffo S 1995, \textit{Phys. Rev. E} \textbf{52} 2361-2374

\bibitem{Sandoz} Antoni M \textit{et al} 1998, \textit{Phys. Rev. E} \textbf{57} 5347–5357

\bibitem{VLatora} Latora V \textit{et al} 2001, \textit{Phys. Rev. E} \textbf{64} 056134

\bibitem{Yamaguchi} Yamaguchi Y Y \textit{et al} 2004, \textit{Physica A} \textbf{337} 36–66

\bibitem{Pluchino2004} Pluchino A \textit{et al} 2004, \textit{Physica A} \textbf{338} 60–67

\bibitem{Leoncini} Leoncini X {et al.} 2009, \textit{EPL} \textbf{86} 20002

\bibitem{Firpo} Firpo M-C 2009, \textit{EPL} \textbf{88} 30010

\bibitem{BouchetGupta2010} Bouchet F \textit{et al} 2010, \textit{Physica A} \textbf{389} 4389 (Special Issue FPSP XII)

\bibitem{Campa} Campa A \textit{et al} 2009, \textit{Phys. Rep.} \textbf{480} 57-159

\bibitem{Yoshida} Yoshida H, 1990 \textit{Phys. Lett. A} \textbf{150} 262-268

\bibitem{Treichler}  Treichler J R 1977, Ph.D. Dissertation, Stanford University

\bibitem{Furedi} F\"{u}redi Z. and Koml\'{o}s J. 1989 \textit{Combinatorica} \textbf{1} 233-241

\bibitem{LB1967} Lynden-Bell D 1967 MNRAS \textbf{136} 101

\bibitem{White} White S D M  in \textit{Gravitational Dynamics}, Proc. 36th Herstmonceux
Conference, ed O. Lahav \textit{et al}, Cambridge University Press,
Cambridge. p121

\bibitem{Severne1986} Severne G and Luwel M 1986, \textit{Astrophys.
Space Sci.} \textbf{122} 299

\bibitem{Mineau1990} Mineau P \textit{et al} 1990, \textit{Astron. Astrophys.} \textbf{228} 344-349

\bibitem{Chavanis2006} Chavanis P H 2006, \textit{Eur. Phys. J. B} \textbf{53} 487

\bibitem{Gabrielli2010} Gabrielli A \textit{et al} 2010, \textit{Phys. Rev. Lett.} \textbf{105} 210602

\bibitem{Barre2010} Barr\'{e} J \textit{et al} 2010, \textit{J. Stat. Mech.} P08002

\bibitem{CampaChavanis2010} Campa A and Chavanis P-H 2010, \textit{J. Stat. Mech.} P06001

\bibitem{Rocha} Rocha Filho T M \textit{et al} 2009, \textit{J. Phys. A: Math. Theor.} \textbf{42}
165001

\bibitem{FLA} Firpo M C \textit{et al} 2006, \textit{Phys. Plasmas} \textbf{13} 122302

\end{thebibliography}

\end{document}